# Meissner Effect and Nonreciprocal Charge Transport in Non-Topological 1T-$CrTe_2$/FeTe Heterostructures

Zi-Jie Yan[1,5], Ying-Ting Chan[2,5], Wei Yuan[1], Annie G. Wang[1], Hemian Yi[1], Zihao Wang[1], Lingjie Zhou[1], Hongtao Rong[1], Deyi Zhuo[1], Ke Wang[3], John Singleton[4], Laurel E. Winter[4], Weida Wu[2*], and Cui-Zu Chang[1,3*]

[1]Department of Physics, The Pennsylvania State University, University Park, PA 16802, USA

[2]Department of Physics and Astronomy, Rutgers University, Piscataway, NJ 08854, USA.

[3]Materials Research Institute, The Pennsylvania State University, University Park, PA 16802, USA

[4]National High Magnetic Field Laboratory, Los Alamos, NM 87544, USA

[5]These authors contributed equally: Zi-Jie Yan and Ying-Ting Chan

[*]Corresponding authors: wdwu@physics.rutgers.edu (W.W.); cxc955@psu.edu (C.-Z. C.)

**Abstract: Interface-induced superconductivity has recently been achieved by stacking a magnetic topological insulator layer on an antiferromagnetic FeTe layer. However, the mechanism driving this emergent superconductivity remains unclear. Here, we employ molecular beam epitaxy to grow a 1T-$CrTe_2$ layer, a two-dimensional ferromagnet with a Curie temperature up to room temperature, on a FeTe layer. These 1T-$CrTe_2$/FeTe heterostructures show superconductivity with a critical temperature of ~12 K. Through magnetic force microscopy measurements, we observe the Meissner effect on the surface of the 1T-$CrTe_2$ layer. Our electrical transport measurements reveal that the 1T-$CrTe_2$/FeTe heterostructures exhibit nonreciprocal charge transport behavior, characterized by a large**

**magneto-chiral anisotropy coefficient. The enhanced nonreciprocal charge transport in 1T-$CrTe_2$/FeTe heterostructures provides a promising platform for exploring the magnetically controllable superconducting diode effect.**

**Main text:** Over the past few decades, emergent phenomena at the interface of two different materials have attracted significant research attention due to their novel and often unexpected physical properties. For example, a two-dimensional (2D) electron gas and superconductivity have been observed at the $LaAlO_3$/$SrTiO_3$ interface[1,2]; the enhanced superconductivity has been achieved in monolayer FeSe/$SrTiO_3$ heterostructures[3-6]. In addition, the interface can naturally induce the proximity effect, enabling one material to obtain the properties of its neighboring material[7-9]. This process can lead to the emergence of various exotic states of matter, including superconductivity, magnetism, and topologically nontrivial phases. Moreover, the formation of heterostructures inherently breaks the inversion symmetry at the interface, potentially giving rise to nonlinear responses[10-15], such as nonreciprocal charge transport in superconducting heterostructures[16-18].

Recently, interface-induced superconductivity and nonreciprocal charge transport have been observed in $(Bi,Sb)_2Te_3$/FeTe heterostructures[19-22], where FeTe is an antiferromagnetic iron chalcogenide that is non-superconducting without element doping[23-25] or tensile stress[26] and $(Bi,Sb)_2Te_3$ is a three-dimensional (3D) topological insulator (TI)[27,28]. Remarkably, this emergent superconductivity persists even after introducing magnetism into the TI layer and coexists with the ferromagnetism or antiferromagnetism in the magnetic TI layer[29,30]. These magnetic TI/FeTe heterostructures provide a promising platform for exploring chiral Majorana physics and developing topological quantum computations. To date, interface-induced superconductivity in FeTe-based heterostructures has been observed exclusively in topological material/FeTe heterostructures[19-22,29,30], with the mechanism behind this superconductivity remains unclear[19-21,31-34]. The large nonreciprocal charge transport in $Bi_2Te_3$/FeTe heterostructures has been attributed to the interplay between the induced superconductivity and the topological Dirac surface states of $Bi_2Te_3$ (Ref.[22]). These observations raise an important question: Is the topological order of the top layer essential for inducing superconductivity in FeTe-based heterostructures? In other words, can superconductivity emerge in nontopological material/FeTe heterostructures, and if so, will the large nonreciprocal charge transport persist therein? To address these questions, it is necessary to find a nontopological material that can replace the TI layer while retaining the interface-induced superconductivity.

1T-$CrTe_2$ is a layered ferromagnet with a trigonal crystal structure, where a Cr layer is sandwiched between two Te layers (Fig. 1a). The Cr atoms exhibit long-range ferromagnetic order along the *c*-axis, and bulk 1T-$CrTe_2$ has a Curie temperature ($T_{Curie}$) above room temperature[35]. Prior studies have shown that intrinsic ferromagnetism persists even when the thickness of 1T-$CrTe_2$ is reduced to a few atomic layers, achieved through either mechanical exfoliation[36] or epitaxial growth[37,38]. Its trivial band structure, ferromagnetic property, and lattice structure similar to the $Bi_2Te_3$ family TI make 1T-$CrTe_2$ an ideal candidate to replace the magnetic TI layer in FeTe-based heterostructures.

In this work, we employ molecular beam epitaxy (MBE) to grow a series of heterostructures by stacking $m$ trilayer (TL) 1T-$CrTe_2$ on $n$ unit-cell (UC) FeTe (Figs. 1a and 1b), referred to as the ($m$, $n$) heterostructures. Through *in situ* reflection high-energy electron diffraction (RHEED) and *ex situ* atomic force microscopy (AFM) measurements, we demonstrate the epitaxy growth of 1T-$CrTe_2$ on FeTe despite their distinct in-plane lattice rotational symmetries. Superconductivity emerges for $m \geq 1$ and $n \geq 4$, with the superconducting temperature ($T_c$) increasing and saturating at ~12 K as both $m$ and $n$ increase. Our magnetic force microscopy (MFM) measurements reveal the Meissner effect in 1T-$CrTe_2$/FeTe heterostructures, confirming the emergence of superconductivity. Further electrical transport measurements show enhanced nonreciprocal charge transport behavior, with the magneto-chiral anisotropy coefficient much larger than previously reported values[22].

First, we characterize the evolution of surface morphology in the FeTe layer before and after the growth of the 1T-$CrTe_2$ layer. The AFM topography image of the 40 UC pristine FeTe layer on a heat-treated $SrTiO_3$(100) substrate reveals square-shaped terraces (Fig. 1c), corresponding to the tetragonal lattice structure of FeTe, with a measured step height of ~0.60 nm between adjacent terraces (Fig. 1c inset). This value is close to the bulk FeTe value of ~0.65 nm (Refs.[39,40]). After depositing 15 TL 1T-$CrTe_2$ layer on 40 UC FeTe, the topography transforms to triangular-shaped terraces, consistent with the trigonal lattice structure of 1T-$CrTe_2$ (Fig. 1d). The step height of the 1T-$CrTe_2$ terraces is ~0.65 nm (Fig. 1d inset), close to its bulk value of ~0.62 nm (Refs.[40,41]). We note that both the 1T-$CrTe_2$ and FeTe layers show sharp and streaky RHEED patterns during the MBE growth (Supplementary Fig. 1), confirming their highly ordered crystal structures.

Next, we conduct RHEED and AFM measurements on a series of ($m$, 40) heterostructures with varying $m$. As $m$ increases, the RHEED patterns from the bottom 40 UC FeTe layer gradually diminish, while the RHEED patterns from the top 1T-$CrTe_2$ become more pronounced, eventually entirely replacing those of FeTe (Supplementary Fig. 1). We observe two sets of RHEED patterns for the 1T-$CrTe_2$ layer (Supplementary Fig. 1f), corresponding to diffractions along $[1\bar{1}00]$ and $[11\bar{2}0]$ directions, respectively. This observation indicates the twin-boundary structure of the 1T-$CrTe_2$ layer, resulting from two possible epitaxial orientations (Fig. 1b). This growth mode arises from the different in-plane rotational symmetries of the 1T-$CrTe_2$ layer (i.e., six-fold) compared to the FeTe layer (i.e., four-fold). Besides RHEED, the AFM images show that the 1T-$CrTe_2$ layer first covers the edges of the square-shaped FeTe terraces (Supplementary Figs. 2a to 2d), gradually extends across the entire films (Supplementary Fig. 2e), ultimately forms the triangular-shapes terraces (Supplementary Fig. 2f). The sharp 1T-$CrTe_2$/FeTe interface is also confirmed by cross-sectional scanning transmission electron microscopy (STEM) and corresponding energy dispersive X-ray spectrometry (EDS) measurements (Fig. 1e and Supplementary Fig. 3). The X-ray diffraction (XRD) spectra and rocking curves of the (15, 40) heterostructure show sharp diffraction peaks from both 1T-$CrTe_2$ and FeTe layers, further validating the high quality of our 1T-$CrTe_2$/FeTe heterostructures (Supplementary Figs. 4 and 5).

Following sample characterization, we perform electrical transport measurements on two series of ($m$, $n$) heterostructures with varying $m$ or $n$. Figure 2a shows the $R_{xx}$-$T$ curves of the ($m$, 40) heterostructures. For $m$ = 0, i.e., the pristine 40 UC FeTe layer, a hump feature is observed at $T \sim 60$ K, corresponding to the paramagnetic-to-antiferromagnetic phase transition of the FeTe layer, known as its *Néel* temperature $T_N$, consistent with prior studies[21,29,30]. For $m$ = 1, a superconducting phase transition is observed, marked by a sudden drop of $R_{xx}$ near $T$ = 7 K, which defines the superconducting onset temperature $T_{c,onset}$. However, a zero-resistance state is absent in the (1, 40) heterostructure down to $T$ =1.7 K, presumably due to the incomplete coverage of the top 1 TL $CrTe_2$ layer, as confirmed by RHEED (Supplementary Fig. 1b) and AFM (Supplementary Fig. 2b) measurements. As $m$ increases, the zero-resistance state appears and persists at higher $T$, accompanied by a sharper superconducting phase transition in the $R_{xx}$-$T$ curves. For $m \geq 3$, the superconducting behavior becomes nearly uniform, with $T_c$ saturating at ~12 K (Figs. 2a and 2c). Figure 2b shows the $R_{xx}$-$T$ curves of the (10, $n$) heterostructures. For $n$ < 4, the (10, $n$) heterostructures exhibit semiconducting behavior from room temperature down to $T$ = 1.7 K, probably due to the inhomogeneity of the FeTe layer. For $n = 4$, $R_{xx}$ shows a sudden drop near $T = 10$ K, indicating a superconducting phase transition. This transition becomes sharper with increasing $n$, and the zero-resistance state is observed for $n \geq 8$. The $T_c$ saturates at ~12 K for $n \geq 15$ (Figs. 2b and 2d). The evolution of superconductivity with increasing $m$ or $n$ indicates that the induced superconducting state becomes spatially inhomogeneous when either the 1T-$CrTe_2$ or FeTe layer is thin, specifically $m < 3$ for $n = 40$ (Fig. 2c) or $n < 15$ for $m = 10$ (Fig. 2d).

To explore the spatial inhomogeneity of the superconducting state, we perform low-temperature MFM experiments to probe the Meissner effect of the induced superconducting state. The Meissner effect is the spontaneous expulsion of the magnetic field in a superconductor, which is the other key characteristic of superconductivity besides the zero-resistance state. The detection of Meissner's response would provide unambiguous evidence of the existence of superconductivity. Although the zero-resistance state has been reported in numerous prior studies on FeTe-based heterostructures, none reported the Meissner effect of the interface-induced superconductivity[19-21,29-34]. MFM detection of the Meissner effect has been reported in various superconductors in either bulk crystals[42-44] or thin films[45,46], where the magnetic tip experiences a repulsive force due to the expulsion of magnetic flux from the sample (Fig. 3a inset). This repulsion results in a positive shift of the resonance frequency of the cantilever, $df \sim \frac{\partial F(z)}{\partial z}$, which can be captured by the $df$-$z$ curve. Here $z$ is the distance between the magnetic tip and the sample surface. Figure 3a shows a concave $df$-$z$ curve measured at a random point on the sample surface at $T = 2.5$ K, i.e., deep inside the superconducting state, clearly demonstrating the Meissner effect of our sample[42-47]. For comparison, the $df$ value remains nearly constant at $T = 17$ K (above superconducting $T_c$), i.e., the normal state.

The spatial distribution of the magnetic response reflects the local phase rigidity in our heterostructures. To investigate the spatial variation of the Meissner repulsion, we take MFM images by scanning the MFM tip at ~250 nm above the sample surface. Several brighter regions appear in the MFM image taken at $T = 8$ K with a 0.2 T magnetic field applied to stabilize the magnetic tip moment (Fig. 3c). This result reveals a nonuniform Meissner's response over a length scale of ~10 μm. We find that the magnetic contrast decreases gradually with increasing temperature and eventually disappears above $T_c$ (~11 K) (Figs. 3c to 3k), which is determined by *in situ* two-terminal resistance $R$ (Fig. 3b). Moreover, the spatial distribution of the Meissner response remains nearly unchanged after thermal cycling across superconducting $T_c$ or under varying applied magnetic fields (Supplementary Figs. 6 and 7). This robustness indicates that the nonuniform distribution is not an extrinsic effect but instead likely associated with static features intrinsic to the 1T-$CrTe_2$/FeTe heterostructures, such as local thickness variations and structural defects.

To confirm that the MFM contrast comes from the inhomogeneity of the Meissner effect, we measure single-point frequency shift $df$ by positioning the MFM tip at locations with bright and dark contrast (Fig. 3e) while cooling the sample through $T_c$ (Fig. 3b). The $df$ signal remains constant above $T_c$, then suddenly rises near $T = 11$ K, marking the onset of the Meissner effect below $T_c$. The observation of the Meissner effect in both locations demonstrates the presence of superconductivity in all locations, and the observed MFM contrast is due to the inhomogeneous Meissner effect. Although spatial inhomogeneity is evident in the Meissner response, the entire film enters the superconducting state for $T < T_c$. The onset of the Meissner repulsion at both the red and green dots occurs at the same temperature within our measurement resolution (Fig. 3b), indicating that the global phase coherence is established across different regions of the 1T-$CrTe_2$/FeTe heterostructure. This finding is consistent with the sharp superconducting transition observed in our electrical transport measurements (Fig. 2a). In other words, the weaker repulsion at the green dot likely reflects a smaller superconducting volume or lower superfluid density than the red dot (Figs. 3b and 3e). Indeed, superconductivity in the regions with low superfluid density would be suppressed in the ultrathin limit, resulting in disconnected superconducting islands, i.e., finite resistance below the mean-field transition temperature. Therefore, the observed inhomogeneity of superfluid density provides a natural mechanism for the thickness dependence of the zero-resistance state. To further examine the spatial distribution of Meissner response in thinner samples, we perform MFM measurements on (1.5, 40) and (10, 10) heterostructures (Supplementary Figs. 8 and 9). Compared to the heterostructures with ($m$, $n$) = (20, 40) and (10, 20) (Fig. 3), we find that reducing either $m$ or $n$ results in a much weaker Meissner response, a shorter inhomogeneity length scale, and a lower $T_c$, indicating the formation of small superconducting patches with weak links. This scenario is consistent with the broader superconducting transition observed in thinner samples (Fig. 2).

The zero-resistance state observed in electrical transport and the Meissner effect detected through MFM unambiguously confirm the interface-induced superconductivity in 1T-$CrTe_2$/FeTe heterostructures. This correlation between the Meissner effect and the zero-resistance state highlights the robustness of the superconducting phase. Observing the Meissner effect in the 1T-$CrTe_2$/FeTe heterostructures provides valuable insights into the nature of the emergent superconductivity. First, despite spatial inhomogeneity, the strong Meissner response observed across the entire scanned area indicates that the emergent superconductivity is not confined to local regions or strictly limited to the $CrTe_2$/FeTe interface but instead extends throughout the film (Fig. 3). Second, the robustness of the Meissner response against thermal cycling and applied magnetic fields points to its origin in static features intrinsic to the $CrTe_2$/FeTe heterostructures, including local strain variations, structural defects, or interface charge transfer (Supplementary Figs. 6 and 7). Moreover, as $T$ decreases, the emerging Meissner effect indicates a strengthening of superconducting coherence, which is expected to enhance the nonreciprocal charge transport behaviors in 1T-$CrTe_2$/FeTe heterostructures.

Next, we investigate the nonreciprocal charge transport in our 1T-$CrTe_2$/FeTe heterostructures. As noted above, the 1T-$CrTe_2$/FeTe interface inherently breaks the inversion symmetry. When an external in-plane magnetic field $\mu_0 H_{\parallel}$ is applied to break the time-reversal symmetry, a nonlinear term is expected to appear in the longitudinal voltage $V_{xx}$ (Refs.[14,16]) (Fig. 4a).

$$V_{xx} = R^{\omega} I + \gamma R^{\omega} \left[ (\mu_0 \overrightarrow{H_{\parallel}} \times \hat{z}) \cdot \vec{I} \right] I \quad (1)$$

Here $\hat{z}$ is the normal direction of the sample plane, $R^{\omega}$ is the first-harmonic longitudinal resistance, and $\gamma$ is the magneto-chiral anisotropy coefficient[16]. The $\gamma$ value quantifies the strength of nonlinear and nonreciprocal charge transport behaviors. When $\mu_0 H_{\parallel}$ is confined within the sample plane, i.e., perpendicular to $\hat{z}$, Equation (1) simplifies to $V_{xx} = R^{\omega} I + \gamma R^{\omega} \mu_0 H_{\parallel} \cos\phi\, I^2$, where $\phi$ is the angle between $\mu_0 H_{\parallel}$ and $\hat{y}$ (Fig. 4a). The second term in this expression accounts for the nonreciprocal charge transport, also known as the second-harmonic signal, due to its quadratic dependence on the applied current $I$. The second harmonic resistance is thus given by $R^{2\omega} = \frac{\gamma R^{\omega} \mu_0 H_{\parallel} \cos\phi I_0}{\sqrt{2}}$, where $I_0$ is the effective value of the applied alternating current (a.c.) (Supplementary Information).

To verify the presence of nonreciprocal charge transport in our 1T-$CrTe_2$/FeTe heterostructures and optimize the second harmonic measurement setup (Methods), we first measure $\phi$-dependent $R^{2\omega}$ and observe a sinusoidal dependence (Fig. 4b inset). For each fixed $\phi$, $R^{2\omega}/R^{\omega}$ is consistently proportional to $\mu_0 H_{\parallel}$ (Fig. 4b). This linear dependence is unaffected by the amplitude and frequency (Supplementary Fig. 10) of the applied a.c. current. These observations align with the above-mentioned theoretical derivations, providing evidence for the emergence of the nonreciprocal charge transport in our 1T-$CrTe_2$/FeTe heterostructures. As $I_0$ increases from 100 μA to 2 mA, the signal-to-noise ratio of the $R^{2\omega}$-$\mu_0 H_{||}$ curves increases (Supplementary Fig. 10a) due to its linear dependence on $I_0$. However, a larger excitation current inevitably suppresses the emergent superconductivity, as suggested by a monotonic increase in $R^{\omega}$ (Supplementary Fig. 10b inset), leading to smaller γ values (Supplementary Fig. 10b). Therefore, we use the excitation current $I_0$ of ~500 μA with a frequency of ~6.447 Hz in our following second-harmonic transport measurements.

Next, we investigate the nonreciprocal charge transport at varying temperatures. For the (10, 20) heterostructure, its Berezinskii-Kosterlitz-Thouless (BKT) transition temperature $T_{BKT}$ is ~11.04 K, determined by fitting its $R_{xx}$-$T$ curve with the Halperin–Nelson formula[48], $R_{xx} \propto \exp(-2b\sqrt{\frac{T_c-T}{T-T_{BKT}}})$, where $b$ is a fitting parameter. For $T < T_{BKT}$, we find that $R^{2\omega}$ vanishes along with $R^{\omega}$, consistent with the zero-resistance state (Supplementary Fig. 11a). For $T_{BKT} < T < T_{c,onset}$, i.e., the superconducting transition regime where the superconducting order parameter appears, the nonreciprocal charge transport arises due to the fluctuating superconducting order parameter (Supplementary Figs. 11b to 11d). In contrast, for $T > T_{c,onset}$, $R^{2\omega}$ becomes negligible again (Supplementary Fig. 11e). The linear relationship between $R^{2\omega}$ and $\mu_0 H_{||}$ is only observed within the superconducting transition regime, i.e., $T_{BKT} < T < T_{c,onset}$. In this regime, electrons begin to form Cooper pairs, shifting the energy scale of the electronic system to change from the Fermi energy ($E_F$ ~ eV) to the superconducting gap size (Δ ~ meV). Therefore, the nonreciprocal charge transport, driven by the spin-orbit interaction and the Zeeman effect, becomes relatively significant compared to the normal states for $T > T_{c,onset}$ (Ref.[16]).

The magneto-chiral anisotropy coefficient $\gamma$ of the (10, 20) heterostructure diverges as $T$ approaches $T_{BKT}$, reaching a maximum value of ~$64.3 \times 10^{-3}$ $T^{-1} \cdot A^{-1}m$ at $T = 11.15$ K (Fig. 4c). This value is an order of magnitude larger than that of the $Bi_2Te_3$/FeTe heterostructures (Ref.[22]). For $Bi_2Te_3$/FeTe heterostructures, the origin of such large nonreciprocal transport is attributed to the Dirac surface states of the $Bi_2Te_3$ layer[22]. However, our second harmonic results suggest that the nonreciprocal transport persists and is even enhanced when the nontopological material 1T-$CrTe_2$ replaces the $Bi_2Te_3$ layer. This observation indicates that the large nonreciprocal charge transport in FeTe-based heterostructures may arise from the FeTe layers near the interface, which become superconducting after the deposition of various top layers. We also perform second harmonic transport measurements on the (5, 40) and (3, 40) heterostructures (Supplementary Fig. 12). Similar behaviors and comparable thickness-independent $\gamma$ values confirm the interfacial origin of the emergent superconductivity in our 1T-$CrTe_2$/FeTe heterostructures.

Finally, we investigate the ferromagnetic properties of our superconducting 1T-$CrTe_2$/FeTe heterostructures. Supplementary Fig. 13a shows the Hall traces of the (10, 20) heterostructure at different temperatures. At $T = 5$ K, i.e., below its $T_c$ (~11.5 K), the Hall resistance $R_{yx}$ remains zero across the entire external magnetic field $\mu_0 H_\perp$ range, consistent with the zero-resistance state. For $T > T_c$, a non-zero $R_{yx}$ with a clear hysteresis loop appears during the $\mu_0 H_\perp$ sweep and persists up to T ~ 200 K, indicating ferromagnetism with $T_{Curie}$ ~ 200 K. Supplementary Fig. 13b summarizes its anomalous Hall resistance $R^{AHE}$ and coercive field $\mu_0 H_c$ as a function of $T$. Similar results are observed in more heterostructures with different $m$ and $n$ (Supplementary Fig. 14). In addition, a sign reversal of $R^{AHE}$ and the topological Hall effect are observed in the pristine 15 TL 1T-$CrTe_2$ film (Supplementary Fig. 15), consistent with the prior studies[49,50]. However, both behaviors are absent in other (*m*, *n*) heterostructures, presumably due to differences in interfacial strain between 1T-$CrTe_2$/FeTe and 1T-$CrTe_2$/$SrTiO_3$(100) interfaces.

To summarize, we use MBE to grow 1T-$CrTe_2$/FeTe heterostructures and observe the emergent superconductivity in these heterostructures. Our MFM measurements reveal the Meissner effect, and our second-harmonic measurements show nonreciprocal charge transport with a large $\gamma$ value. The observations of the Meissner effect and nonreciprocal charge transport in our 1T-$CrTe_2$/FeTe heterostructures confirm that the topological surface states of the top layer are not a prerequisite for creating superconductivity in FeTe-based heterostructures. Our results indicate that the FeTe layer near the interface is most likely responsible for the emergent superconductivity. We hypothesize that the Te element may be crucial for the formation of the superconductivity due to its consistent presence in the top layer of various FeTe-based superconducting heterostructures[19-22,29-34]. The 1T-$CrTe_2$/FeTe heterostructures are also promising candidates for exploring magnetically controllable superconducting diode effect, even at zero magnetic field.

**Methods**

**MBE growth**

The 1T-$CrTe_2$/FeTe heterostructures used in this work are grown on insulating 0.5 mm thick $SrTiO_3$(100) substrates in a commercial MBE chamber (ScientaOmicron Lab10) with the vacuum better than ~$2 \times 10^{-10}$ mbar. The $SrTiO_3$(100) substrates are first soaked in ~80 °C deionized water for ~2 hours and then diluted hydrochloric acid solution (~4.5% w/w) for ~2 hours. Next, these $SrTiO_3$(100) substrates are annealed in a tube furnace with flowing high-purity oxygen gas at ~974 °C for ~3 hours. These heat treatments make the $SrTiO_3$(100) surface passivated, suitable for the MBE growth of 1T-$CrTe_2$/FeTe heterostructures. The heat-treated $SrTiO_3$ (100) substrates are loaded into the MBE chamber and outgassed at ~600 °C for ~1 hour before the MBE growth. High-purity Fe (99.995%), Te (99.9999%), and Cr (99.999%) are evaporated from Knudsen effusion cells. The growth temperature is ~340 °C for FeTe and ~ 300 °C for 1T-$CrTe_2$. The growth rate is ~0.3 UC per minute for the FeTe layer and ~0.25 TL per minute for the 1T-$CrTe_2$ layer, calibrated by AFM and STEM measurements. No capping layer is involved in our measurements.

**XRD measurements**

The high-resolution XRD measurements are carried out at room temperature using a Malvern Panalytical X'Pert3 MRD with Cu-$K_{\alpha 1}$ radiation (wavelength λ~1.5405980 Å).

**HAADF-STEM measurements**

The Aberration-corrected HAADF-STEM measurements are performed on an FEI Titan$^3$ G2 operating parameters at an accelerating voltage of ~300 kV, with a probe convergence angle of ~30 mrad, a probe current of ~70 pA, and HAADF detector angles of 52~253 mrad. More STEM images are shown in Supplementary Figs. 3 and 21a.

**Electrical transport measurements**

The 1T-$CrTe_2$/FeTe heterostructures grown on 2 mm × 10 mm $SrTiO_3$ (100) substrates are scratched into a Hall bar geometry using a computer-controlled probe station (Supplementary Fig. 21b). The effective area of the Hall bar device is ~ 1 mm × 0.5 mm. The electrical contacts are made by pressing indium spheres on the Hall bar. The electrical transport measurements are conducted using two Physical Property Measurement Systems (PPMS, Quantum Design DynaCool, 1.7 K, 9 T/14 T). The excitation current is ~1 μA for $R_{xx}$-$T$ measurements and ~100 μA for Hall measurements. The magneto-transport results are symmetrized or anti-symmetrized to eliminate mutual pick-up caused by slight geometrical misalignment of the electrodes. Electrical transport measurements under high magnetic fields (>14 T) are conducted in a capacitor-driven 65 T pulsed magnet at the National High Magnetic Field Laboratory (NHMFL) in Los Alamos. More transport results are shown in Supplementary Figs. 10 to 20.

**Nonreciprocal charge transport measurements**

The nonreciprocal charge transport measurements are performed using a PPMS (Quantum Design DynaCool, 1.7 K, 9 T) with a single-axis horizontal rotator module (Supplementary Fig. 22), enabling in-plane angular rotation of the sample with a ~0.1° precision. The magnetic field is applied within the rotational plane, and thus remains strictly in-plane relative to the sample surface during rotation. To minimize canting or misalignment, the sample surface is carefully aligned to the rotational plane. A Keithley 6221 source meter injects an a.c. current through the samples, while the first- and second-harmonic voltages are measured by SRS860 lock-in amplifiers. Unless otherwise specified, the excitation current is fixed at ~500 μA with a frequency of ~6.447 Hz for all measurements, and the external in-plane magnetic field is applied perpendicular to the excitation current. The second-harmonic resistance has been anti-symmetrized to eliminate pick-up signals from the first-harmonic resistance. The $\gamma$ values are obtained by conducting linear regressions to the $R^{2\omega}/R^{\omega}$-$\mu_0 H_{\parallel}$ curves.

**MFM measurements**

The MFM measurements are performed with a homebuilt Helium-3 AFM system using commercial piezoresistive cantilevers (spring constant $k \sim 3$ N/m and resonant frequency $f_0 \sim 44$ kHz). The MFM tips are coated with a nominally 100 nm thick Co layer using magnetron sputtering. The MFM results are extracted by Nanonis SPM Controllers (SPECS) with a phase-locked loop. The MFM signal, the resonant frequency shift of the cantilever, is proportional to the out-of-plane stray field gradient generated by the sample. MFM images are acquired on a scanning plane 200~250 nm above the surface using the constant-height mode. For balanced visualization of the spatial distribution of the MFM signal at each temperature (i.e., the frequency shift *df*), the color scale is centered at the average MFM signal of each image, with the full color scale mapping to the relative variation of the MFM signal in each image. *In situ* two-terminal resistance is measured together with the MFM measurements. To eliminate the electrical contact and lead wire resistance, the residual resistance at $T = 2$ K has been subtracted from the $R$-$T$ curve (Fig. 3b). The MFM contrast in Fig. 3b is three times the RMS value of each MFM image, with the value at $T =$ 14 K used as an offset to remove the finite value contributed by noise and background.

**Acknowledgments:** This work is primarily supported by the ONR Award (N000142412133), including the MBE growth and PPMS (9T) measurements. The AFM and XRD measurements are partially supported by the DOE grant (DE-SC0023113). The STEM and PPMS (14T) measurements are partially supported by Penn State MRSEC for Nanoscale Science (DMR-2011839). The MFM measurements are supported by the DOE grant (DE-SC0018153). C. -Z. C. acknowledges the support from the Gordon and Betty Moore Foundation's EPiQS Initiative (Grant GBMF9063 to C. -Z. C.). Work done at NHMFL is supported by NSF (DMR-2128556) (J. S. and L. E. W.) and the State of Florida.

**Author contributions:** C. -Z. C. and W. W. conceived and designed the experiment. Z. -J. Y., and W. Y. performed the MBE growth. Z. -J. Y., W. Y., A. G. W., Z. W., L. -J. Z., H. R., and D. Z. performed electrical transport measurements. Y.-T.C. and W. W. performed the MFM measurements. Z. -J. Y. performed AFM and XRD measurements. J. S., L. E. W., H. Y., and Z. -J. Y. performed the electrical transport measurements under high magnetic fields. K. W., H. Y., and Z. -J. Y. carried out the STEM measurements. Z. -J. Y., Y. -T. C., W. W., and C. -Z. C. analyzed the data and wrote the manuscript with input from all authors.

**Competing interests:** The authors declare no competing interests.

**Data availability:** The datasets generated during and/or analyzed during this study are available from the corresponding author upon reasonable request.

**Figures and figure captions:**

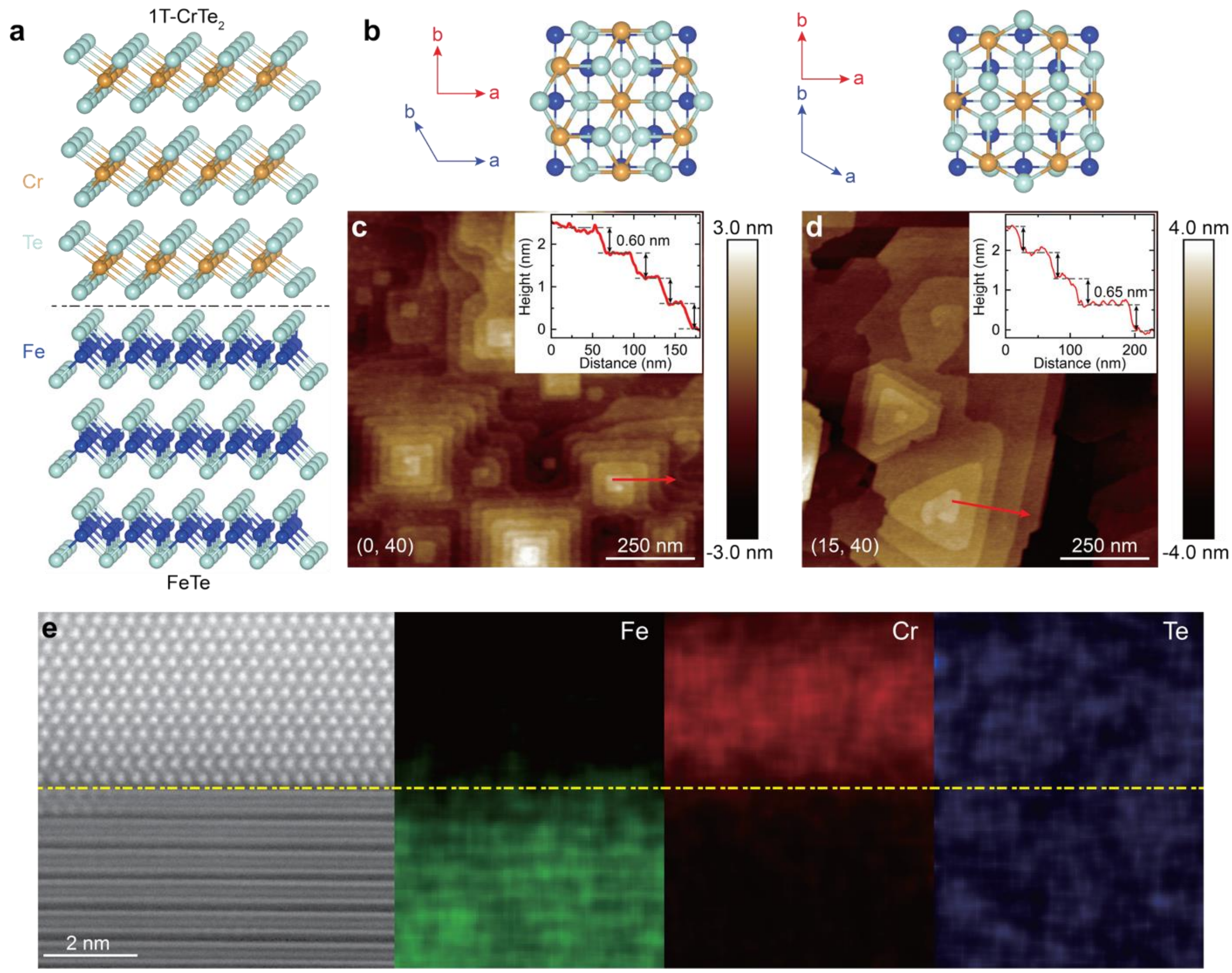

**Fig. 1| MBE-grown 1T-$CrTe_2$/FeTe heterostructures on $SrTiO_3$(100).** **a**, Schematic lattice structure of the1T-$CrTe_2$/FeTe heterostructure. **b**, Two possible stacking orientations of 1T-$CrTe_2$ on FeTe. The blue (red) arrows represent the crystal axes of the top 1T-$CrTe_2$ (bottom FeTe) layer. **c, d**, 1 μm × 1 μm AFM images of 40 UC FeTe/$SrTiO_3$(100) (**c**) and 15 TL $CrTe_2$/40 UC FeTe/$SrTiO_3$(100) (**d**). Insets: the height profiles along the red arrows. **e**, Cross-sectional STEM image of the 15 TL $CrTe_2$/40 UC FeTe heterostructure with EDS mappings of Fe, Cr, and Te elements, respectively. The yellow dashed line indicates the sharp interface between 1T-$CrTe_2$ and FeTe.

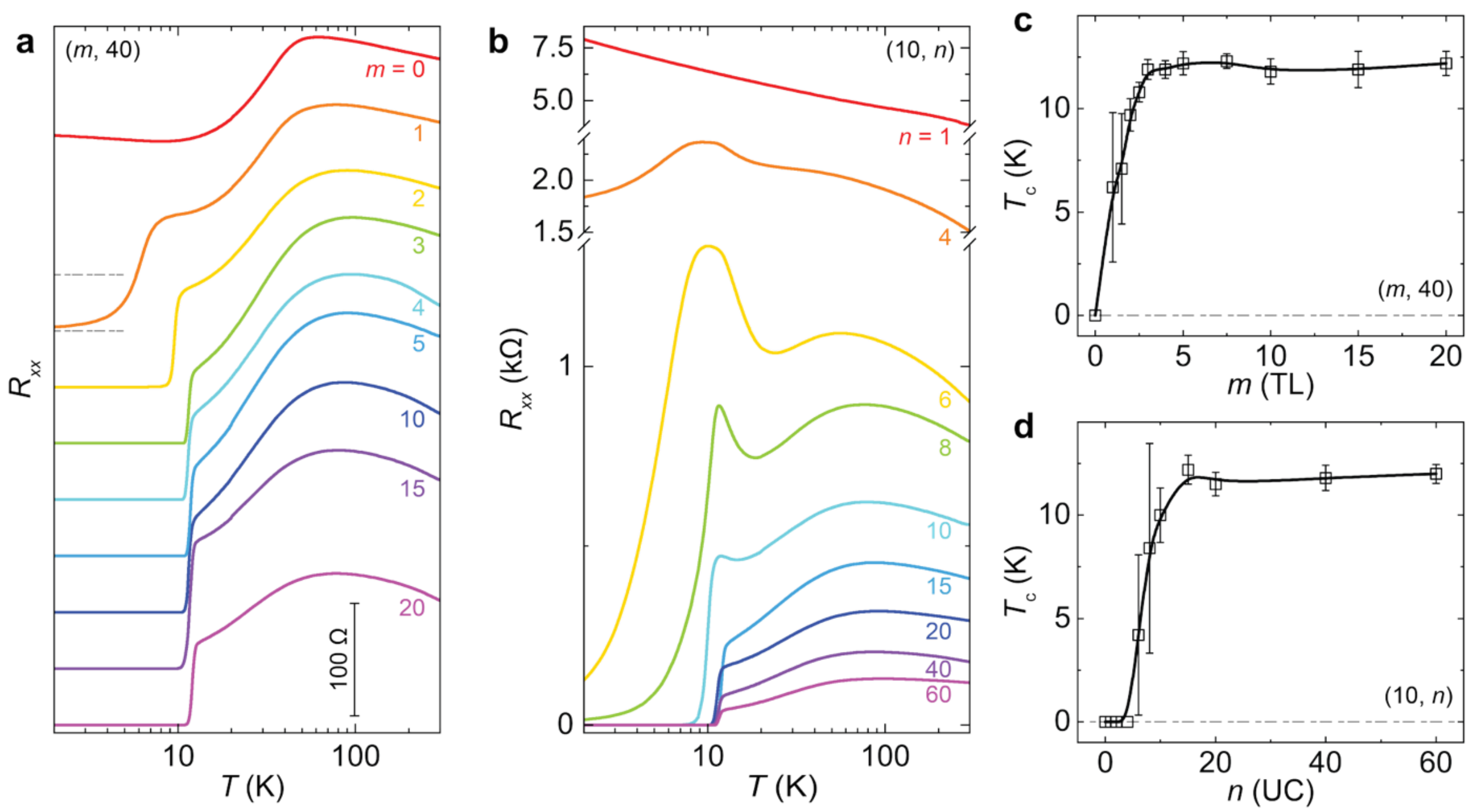

**Fig. 2| Interface-induced superconductivity in 1T-$CrTe_2$/FeTe heterostructures. a**, $T$ dependence of the sheet longitudinal resistance, $R_{xx}$, of the ($m$, 40) heterostructures with $0 \leq m \leq 20$. Each curve is shifted by 50 Ω. The two horizontal dashed lines indicate the zero resistance of the $m = 0$ and $m = 1$ samples. **b**, $T$ dependence of $R_{xx}$ of the (10, $n$) heterostructures with $1 \leq n \leq 60$. **c,** $m$ dependent superconducting temperature $T_c$ of the ($m$, 40) heterostructures. **d,** $n$ dependent $T_c$ of the (10, $n$) heterostructures. The $T_c$ value is the temperature at which $R_{xx}$ drops to 50% of its normal state resistance. The error bar of each sample is determined from the value difference between $T_{c,onset}$ and $T_{c,0}$.

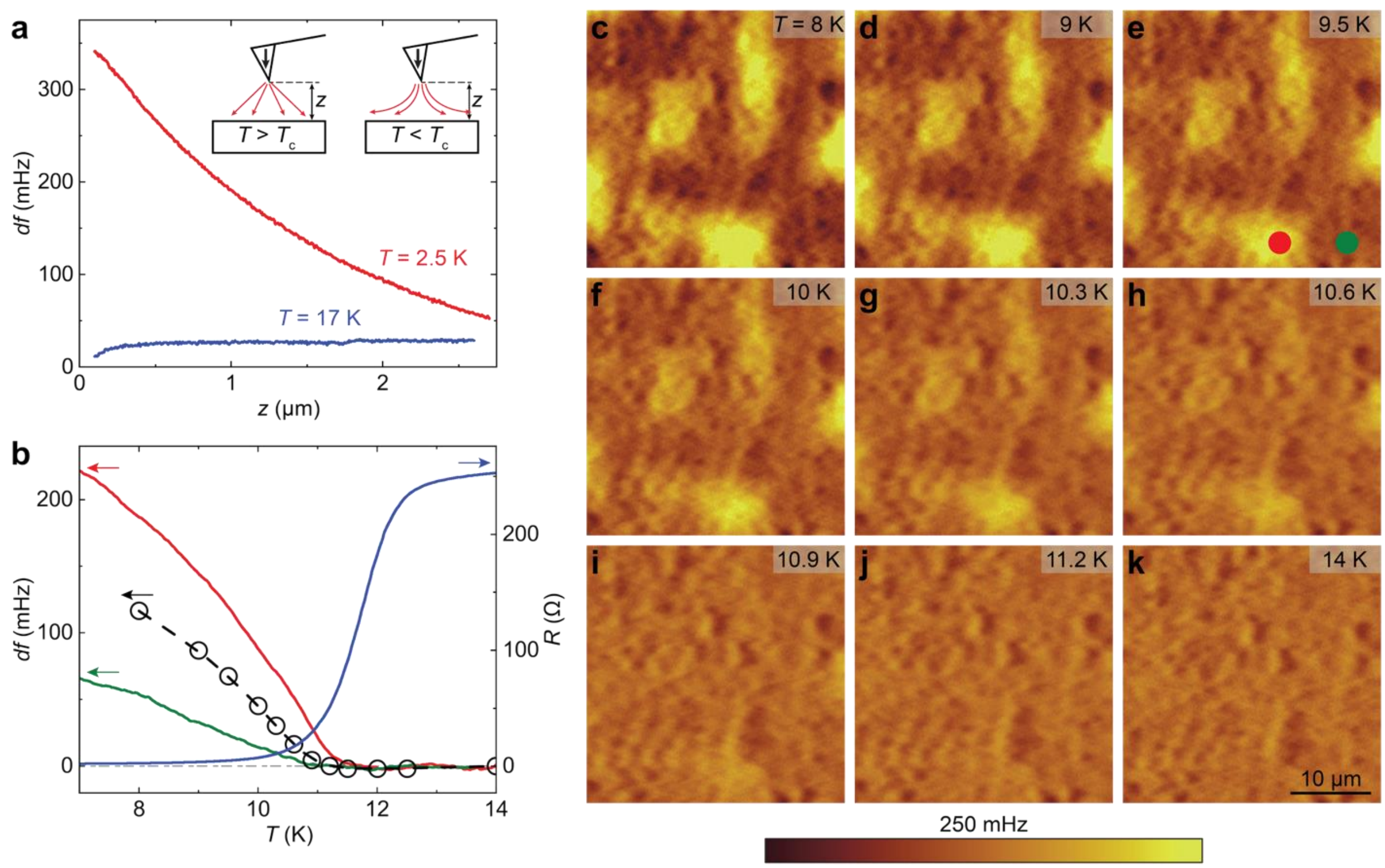

**Fig. 3| Meissner effect in 1T-$CrTe_2$/FeTe heterostructures. a**, $z$-dependence of MFM frequency shift $df$ measured on the (10, 20) heterostructure at $T$ = 2.5 K (red) and $T$ = 17 K (blue). Inset: Schematic of the Meissner effect experienced by the MFM tip. For $T > T_c$, i.e., the normal state, the stray field of the magnetic tip can penetrate the heterostructure. However, for $T < T_c$, i.e., the superconducting state, the Meissner effect partially expels the stray field and generates a repulsive force on the magnetic tip. **b**, $T$ dependence of $df$ measured on the red and green dots in (**e**) (red and green curves), MFM contrasts (black circles), and *in situ* two-terminal resistance $R$ (blue curve) of the (20, 40) heterostructure. The MFM contrasts are defined as three times of the root mean square (RMS) value of each MFM image (Methods). **c-k**, MFM images of the (20, 40) heterostructure measured at $T$ = 8 K (**c**), $T$ = 9 K (**d**), $T$ = 9.5 K (**e**), $T$ = 10 K (**f**), $T$ = 10.3 K (**g**), $T$ = 10.6 K (**h**), $T$ = 10.9 K (**i**), $T$ = 11.2 K (**j**), and $T$ = 14 K (**k**). During MFM measurements, the magnetic tip is ~250 nm above the sample surface, and an external magnetic field of ~0.2 T is applied.

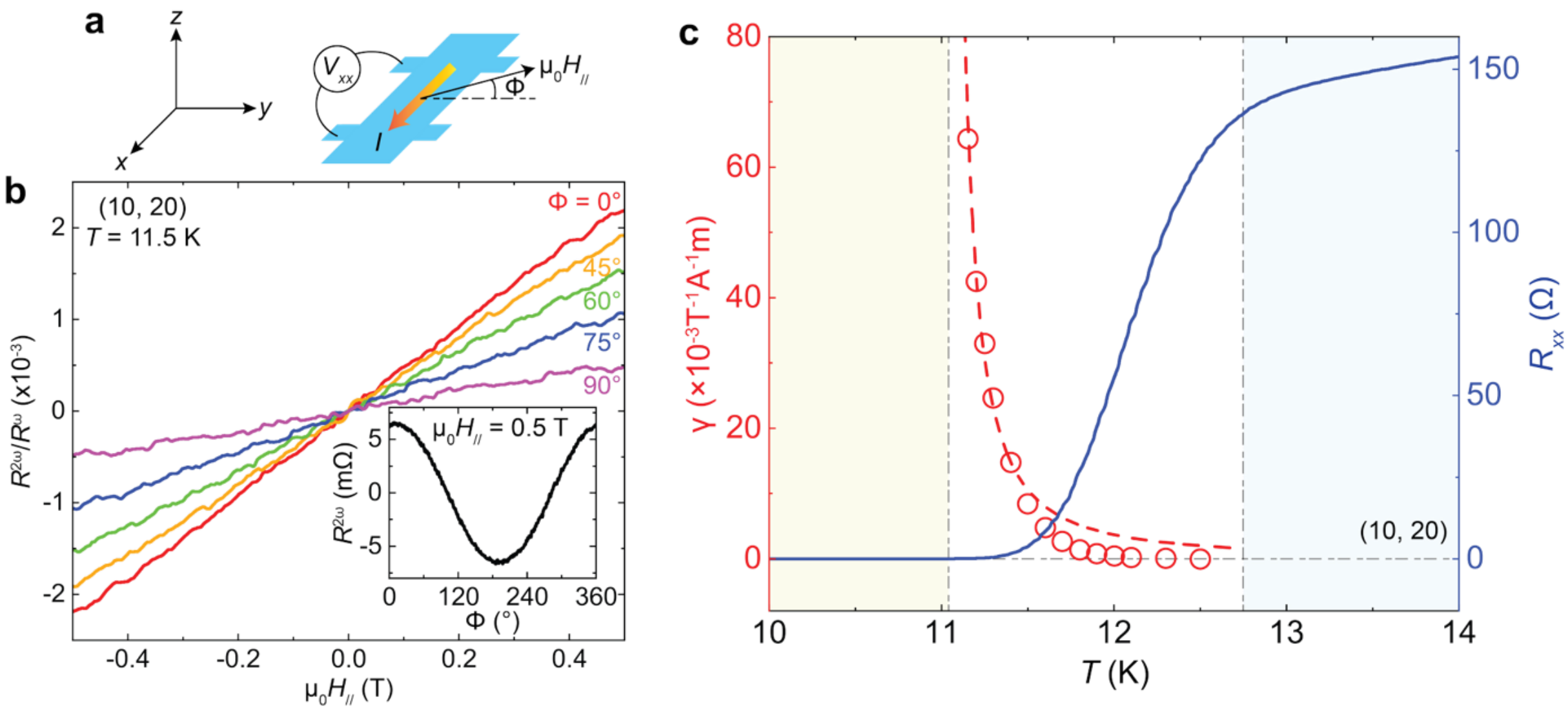

**Fig. 4| Nonreciprocal charge transport in the (10, 20) heterostructure. a**, Schematic of the nonreciprocal charge transport measurement setup. The angle $\phi$ is defined between the in-plane magnetic field $\mu_0H_{||}$ and the $y$-axis ($\hat{y}$) direction. **b**, $\mu_0H_{||}$ dependence of second-harmonic response (i.e., $R^{2\omega}/R^{\omega}$) measured under different $\phi$ and at $T$ = 11.5 K. Inset: $\phi$ dependence of the second-harmonic resistance $R^{2\omega}$ measured at $T$ = 11.5 K under a fixed $\mu_0H_{||}$ = 0.5 T. **c**, $T$ dependence of magneto-chiral anisotropy coefficient γ (red circles) and the sheet longitudinal resistance $R_{xx}$ (blue curve). The red dashed curve is fitted by the formula $\gamma = \beta(T - T_{BKT})^{-1.5}$, where $T_{BKT}$ is the BKT superconducting transition temperature, and β is a fitting coefficient. The large nonreciprocal charge transport occurs only during the superconducting transition regime.